\documentclass{mn2e}
\usepackage{epsfig}
\usepackage{epsf}
\usepackage{graphicx}

\title[Light Curve of Eta Car]{A Revised Historical Light Curve of Eta
  Carinae and the Timing of Close Periastron Encounters}

\author[Smith \& Frew]{Nathan Smith$^1$\thanks{Email:
    nathans@astro.berkeley.edu} and David J. Frew$^2$ \\ $^1$Steward
  Observatory, University of Arizona, 933 North Cherry Avenue, Tucson,
  AZ 85721, USA \\ $^2$Department of Physics and Astronomy, Macquarie
  University, North Ryde, NSW 2109, Australia}

\begin{document}
\date{Accepted 0000, Received 0000, in original form 0000}
\pagerange{\pageref{firstpage}--\pageref{lastpage}} \pubyear{2002}
\def\arcdeg{\degr}
\maketitle
\label{firstpage}

\begin{abstract}

  The historical light curve of the 19th century ``Great Eruption'' of
  $\eta$~Carinae provides a striking record of the violent instabilies
  encountered by the most massive stars.  In this paper we report and
  analyze newly uncovered historical estimates of the visual
  brightness of $\eta$~Car during its eruption, and we correct some
  mistakes in the original record.  The revised historical light curve
  looks substantially different from previous accounts: it shows two
  brief precursor eruptions in 1838 and 1843 that resemble modern
  supernova impostors, while the final brightening in December 1844
  marks the time when $\eta$~Car reached its peak brightness.  We
  consider the timing of brightening events as they pertain to the
  putative binary system in $\eta$~Car: (1) The brief 1838 and 1843
  events rose to peak brightness within weeks of periastron passages
  if the pre-1845 orbital period is $\sim$5\% shorter than at present
  due to the mass loss of the eruption.  Each event lasted only
  $\sim$100 days.  (2) The main brightening at the end of 1844 has no
  conceivable association with periastron, beginning suddenly more
  than 1.5 yr {\it after} periastron.  It lasted $\sim$10 yr, with no
  obvious influence of periastron encounters during that time.  (3)
  The 1890 eruption {\it began} to brighten at periastron, but took
  over 1 yr to reach maximum brightness and remained there for almost
  10 yr.  A second periastron passage midway through the 1890 eruption
  had no visible effect.  While the evidence for a link between
  periastron encounters and the two brief precursor events is
  compelling, the differences between the three cases above make it
  difficult to explain all three phenomena with the same mechanism.

\end{abstract}


\section{INTRODUCTION}

Among massive stars, the enigmatic object $\eta$~Carinae is
simultaneously our most scrutinized case study and still among the
most mysterious (Davidson \& Humphreys 1997).  Its bipolar Homunculus
Nebula provides proof that massive stars can eject more than 10
$M_{\odot}$ (Smith et al.\ 2003b) in a single eruptive event and
survive, while the present-day star and its putative binary companion
present a number of enduring challenges.

The central mystery concerning $\eta$~Car is the cause of its
spectacular ``Great Eruption'' in the mid-19th century (Davidson \&
Humphreys 1997), when it displayed erratic variability and briefly
became the second brightest star in the sky despite its distance of
$\sim$2.3 kpc (Smith 2006).  Observing $\eta$~Carinae at the Cape of
Good Hope in the early to mid-19th century, J.F.W.\ Herschel first
described the ``sudden flashes and relapses'' of $\eta$~Argus, as it
was called at the time, and remarked that this star was ``fitfully
variable to an astonishing extent'' (Herschel 1847). At times it
rivaled Sirius and Canopus in brightness, but with an orange-red
colour.  Innes (1903) compiled a list of known 19th-century
observations and published the familiar lightcurve that has been often
reproduced and supplemented by modern observations.  The lightcurve
was updated and corrected for scale errors by Frew (2004).

The complex Homunculus Nebula surrounding $\eta$~Car is a prototypical
bipolar nebula, made famous in spectacular images made with the {\it
  Hubble Space Telescope} ({\it HST}) (e.g., Morse et al.\ 1998).  It
had long been suspected that the Homunculus originated from the Great
Eruption (Gaviola 1950; Ringuelet 1958; Gehrz \& Ney 1972), and
proper-motion measurements of the expanding nebula later confirmed
this, with estimated ejection dates of 1841 (Currie et al.\ 1996),
1844 (Smith \& Gehrz 1998), and 1846--1848 (Morse et al.\ 2001).  That
the historical brightening event was observed and that we can now
study its expanding ejecta make $\eta$~Car uniquely valuable.

Multiple eruptive episodes in $\eta$~Car point to an enduring phase of
instability, marked by repeated sequences of outburst and recovery.
In addition to the multiple peaks during the Great Eruption that we
discuss in this paper, the star brightened again around 1890 when it
ejected another bipolar nebula called the Little Homunculus (Ishibashi
et al.\ 2003; Smith 2002, 2005).  Additional nebulosity outside the
Homunculus suggests major ancient eruptions 500--2000 yr ago (Walborn
\& Blanco 1988; Walborn et al.\ 1978; Smith \& Morse 2004; Smith et
al.\ 2005).  The star has also been brightening in a non-steady way in
modern times, with a jump in the 1940s (de Vaucouleurs \& Eggen 1952)
and again in the late 1990s (Davidson et al.\ 1999).

As our best studied example, $\eta$ Car serves as the prototype for a
class of transient sources known variously as giant LBV eruptions,
Type V supernovae (SNe), SN impostors, or $\eta$ Car analogs, which
are thought to represent non-terminal eruptions of massive stars.
Smith et al.\ (2010) recently provided a comprehensive study and
review of this class of objects and related transient sources.
Although $\eta$ Car is often held as the prototype for this class, it
is hardly a typical case.  Its multiple peaks and long duration are
unusual, although not unique (Smith et al.\ 2010).  Unlike all
extragalactic SN impostors, its nebula can be studied in detail, and
linking clues from the spatially resolved ejecta to the timing of
brightening events is a critical piece of the puzzle.  As such, here
we aim to provide the definitive historical light curve of $\eta$ Car
after reviewing all available historical documentation.  A detailed
discussion of each observation in the historical light curve of
$\eta$~Carinae known at that time was given by Frew (2004).  In this
paper, we collect and present 51 previously unpublished estimates of
the brightness of $\eta$~Car from C.P.\ Smyth and Thomas Maclear in
the 1840s.  In combination with the data from Frew (2004), this
archival data set of new measurements has allowed the first detailed
look at the photometric behavior of $\eta$~Carinae during critical
points in its Great Eruption of 1837--1858.  Including the new
archival data, the character of the light curve is different from
previous reports concerning the period of time centered around the
peak of the Great Eruption, as we discuss below.

\section{OBSERVATIONS}

\subsection{New Archival Data}

The recent digital publication of the Royal Astronomical Society's
Herschel Archives (Hoskin 2005) has been a boon to historians of
astronomy.  An examination of this resource has revealed an important
new data set of $\eta$~Carinae's brightness in 1842--43, compiled by
Charles Piazzi Smyth, which has now been fully reduced.  This data set
was not summarised by Herschel (1847), and it was not available at the
time Frew (2004) compiled the historical light curve for $\eta$~Car.
We also took the opportunity to examine the Royal Society Archives
that were available on microfilm from the University Publications of
America (1990).  We perused more than 100 letters from Thomas Maclear
and C.P.\ Smyth (both were observers at the Cape of Good Hope) to Sir
John Herschel between 1838 and 1865, in order to search for any
additional unpublished observations.  The index of Crowe, Dyck \&
Kevin (1998) permitted an efficient search through the archival
letters.

\begin{figure*}\begin{center}
\includegraphics[width=4.8in]{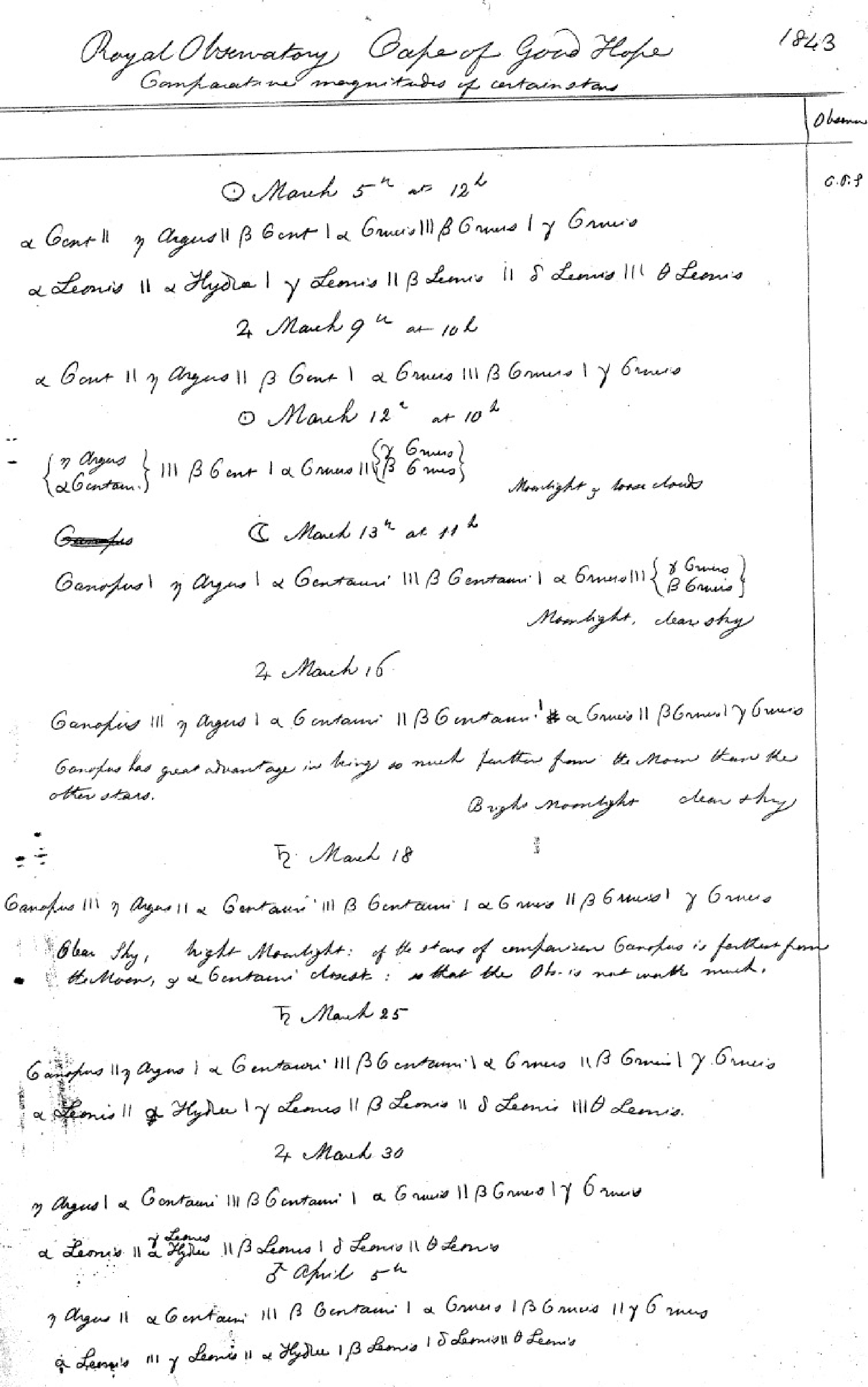}
\end{center}
\caption{Example presenting the original source material, showing the
  list of comparisons by C.P. Smyth (1843).  Reproduced from the
  Herschel Archives of the Royal Astronomical
  Society.}\label{fig:smyth}
\end{figure*}

These new observations have cleared up some ambiguities and
inconsistencies in the data summarised by Herschel (1847).  The
discrepancies pointed out by M\"uller \& Hartwig (1918) have now been
clarified after examining these original archival letters.  We have
reproduced a page from Smyth's (1843) manuscript as
Figure~\ref{fig:smyth} to illustrate the nature and scope of the
source material.

\begin{table}\begin{center}\begin{minipage}{3.1in}
      \caption{Observations of $\eta$ Argus by Smyth and
        Maclear}\scriptsize
\begin{tabular}{@{}lccc}\hline\hline
Observer	&Date		&m$_v$	&err	\\
		&(year)		&(mag)	&(mag)	\\ \hline	
Maclear, T.	&1842.213	&1.0	&0.4	\\
Maclear, T.	&1842.995	&0.7	&0.2	\\	
Maclear, T.	&1842.997	&0.2	&0.2	\\	
Maclear, T.	&1843.000	&0.3	&0.2	\\	
Maclear, T.	&1843.005	&0.3	&0.2	\\	
Maclear, T.	&1843.022	&0.3	&0.2	\\	
Maclear, T.	&1843.027	&0.3	&0.2	\\	
Maclear, T.	&1843.030	&0.3	&0.2	\\	
Maclear, T.	&1843.033	&0.3	&0.2	\\	
Maclear, T.	&1843.036	&0.3	&0.2	\\	
Maclear, T.	&1843.038	&0.3	&0.2	\\	
Smyth, C.P.	&1843.044	&0.4	&0.2	\\	
Smyth, C.P.	&1843.047	&0.4	&0.2	\\
Smyth, C.P.	&1843.049	&0.6	&0.2	\\
Smyth, C.P.	&1843.079	&0.4	&0.2	\\
Smyth, C.P.	&1843.088	&0.4	&0.2	\\
Smyth, C.P.	&1843.112	&0.4	&0.2	\\
Smyth, C.P.	&1843.137	&0.3	&0.2	\\
Smyth, C.P.	&1843.164	&0.3	&0.2	\\
Smyth, C.P.	&1843.170	&0.3	&0.2	\\
Smyth, C.P.	&1843.175	&0.2	&0.2	\\
Smyth, C.P.	&1843.186	&0.2	&0.2	\\
Maclear, T.	&1843.192	&-0.8	&0.3	\\
Smyth, C.P.	&1843.195	&-0.3	&0.2	\\
Smyth, C.P.	&1843.197	&-0.5	&0.2	\\
Maclear, T.	&1843.200	&-0.8	&0.3	\\
Smyth, C.P.	&1843.205	&-0.4	&0.2	\\
Maclear, T.	&1843.208	&-0.5	&0.3	\\
Maclear, T.	&1843.211	&-0.5	&0.3	\\
Smyth, C.P.	&1843.211	&-0.5	&0.2	\\
Maclear, T.	&1843.214	&-0.5	&0.3	\\
Maclear, T.	&1843.227	&-0.4	&0.3	\\
Smyth, C.P.	&1843.230	&-0.4	&0.2	\\
Maclear, T.	&1843.238	&-0.3	&0.3	\\
Smyth, C.P.	&1843.244	&-0.5	&0.3	\\
Leps		&1843.249	&-0.7	&0.3	\\
Smyth, C.P.	&1843.260	&-0.8	&0.3	\\
Smyth, C.P.	&1843.282	&-0.5	&0.3	\\
Smyth, C.P.	&1843.299	&-0.8	&0.3	\\
Smyth, C.P.	&1843.323	&-0.5	&0.3	\\
Smyth, C.P.	&1843.326	&-0.3	&0.2	\\
Smyth, C.P.	&1843.359	&-0.3	&0.2	\\
Smyth, C.P.	&1843.389	&-0.1	&0.2	\\
Smyth, C.P.	&1843.400	&-0.3	&0.2	\\
Smyth, C.P.	&1843.411	&-0.3	&0.2	\\	
Maclear, T.	&1844.05	&0.2	&0.4	\\	
Maclear, T.	&1844.71	&0.2	&0.3	\\	
Smyth, C.P.	&1844.92	&-1.0	&0.3	\\
Maclear, T.	&1844.96	&-1.0	&0.3	\\
Smyth, C.P.	&1845.00	&-1.0	&0.3	\\
Smyth, C.P.	&1845.79	&-0.6	&0.3	\\
\hline
\end{tabular}\label{tab:smyth}
\end{minipage}\end{center}
\end{table}

\subsection{Reduction Procedure}

Following the procedure used in Frew (2004), all brightness estimates
were reduced to the photopic visual system, with the zero point
equivalent to Johnson $V$ for an A0 star.  Almost all visual observers
naturally use photopic (foveal) vision, with an effective wavelength,
$\lambda_{\rm eff}$ of 5600 \AA, only slightly redward of Johnson $V$
($\lambda_{\rm eff}$ of 5450 \AA).  We note that scotopic (peripheral
or rod) vision ($\lambda_{\rm eff}$ $\simeq$ 5100 \AA) is considerably
bluer than $V$, but is almost never used for stellar brightness
estimation; see Schaefer (1996b) and Frew (2004) for a fuller
discussion.

As before, $V$ magnitudes for the comparison stars were taken from the
Lausanne photometric database (Mermilliod, Mermilliod, \& Hauck 1997).
No corrections for differential extinction were applied, as any factor
is likely to be smaller than the adopted uncertainties of the visual
magnitudes, derived from the interpolative method used.

The brightness descriptions of Smyth (1843) are in raw form and are
equivalent to the traditional Argelander step method, where the
difference in brightness of stars along a defined sequence is
estimated. Smyth (1843) described his observing method as follows:
``The stars are here put down in their order of lustre as estimated by
the naked eye.  The vertical strokes are intended to show the supposed
number of grades between any two.''

An extract from his manuscript is reproduced here as
Figure~\ref{fig:smyth}, and we use his data to determine the visual
magnitude of $\eta$~Carinae by interpolation.  We illustrate our
reduction method using Smyth's observations for March 18, 1843; the
values in parentheses are the grades in brightness estimated by Smyth:

\smallskip

{\small Canopus (3) $\eta$ Argus (2) $\alpha$ Centauri (3) 
$\beta$ Centauri (1) $\alpha$ Crucis (2) $\beta$ Crucis (1) 
$\gamma$ Crucis }

\smallskip

Utilising the comparison star magnitudes given in Table~3 of Frew
(2004), it can be seen that $\eta$~Carinae had $m_{\rm V}$ = $-$0.5
$\pm$0.2 on this date.  We note that on this night, the magnitude of
each grade or step was not constant along the sequence, ranging from
0.09 mag between Canopus and $\alpha$~Centauri to 0.37 mag between
$\beta$ and $\gamma$~Crucis.  This is in fact typical of each night's
data.  Using all of the data from Smyth's manuscript leads us to adopt
a mean step value of $\Delta$m = 0.24 $\pm$0.13 mag ($n$ = 160).  On
some nights (e.g., April 19, 1843) $\eta$~Carinae was brighter than
the brightest comparison star, so the derived $\Delta$m value has been
used to determine the magnitude of $\eta$~Carinae from extrapolation,
with a larger uncertainty of 0.3 mag adopted as a result (see also
Frew 2004).

Our derived visual magnitudes are denoted throughout by $m_{\rm V}$,
and realistic uncertainties have been determined for each data point.
For the majority of observers there will be only a small colour term
between the visual system and Johnson $V$ for most naked-eye stars,
but the difference between the $m_{\rm V}$ and $V$ systems for an
emission-line star like $\eta$~Carinae might be substantial.  We do
not currently have enough information to quantify the effects of the
emission-line spectrum during the eruption, but we note that any error
is likely to be less than the generous adopted uncertainty of
$\pm$0.2--0.5 mag.

The observations of Thomas Maclear are reduced in a similar way
(Maclear 1842, 1843, 1844a,b).  Maclear also used a step method, but
his descriptions are more verbose and less precise.  An example from
March 24, 1843 is typical (Maclear 1843):

\smallskip

{\small ``Decidedly not so brilliant as Canopus, brighter than
  $\alpha^{1,2}$ Centauri.'' }

\smallskip

From this description, the concluded magnitude is m$_{V}$ = $-$0.5
$\pm$0.2.  Another example is Maclear's observation of Mar 19, 1842.
Maclear (1842) wrote:

\smallskip

{\small ``...it was considerably less than Rigel, less than $\alpha$
  Crucis \& much greater than $\alpha$ Hydrae.'' }

\smallskip

\begin{figure*}\begin{center}
\includegraphics[width=5.7in]{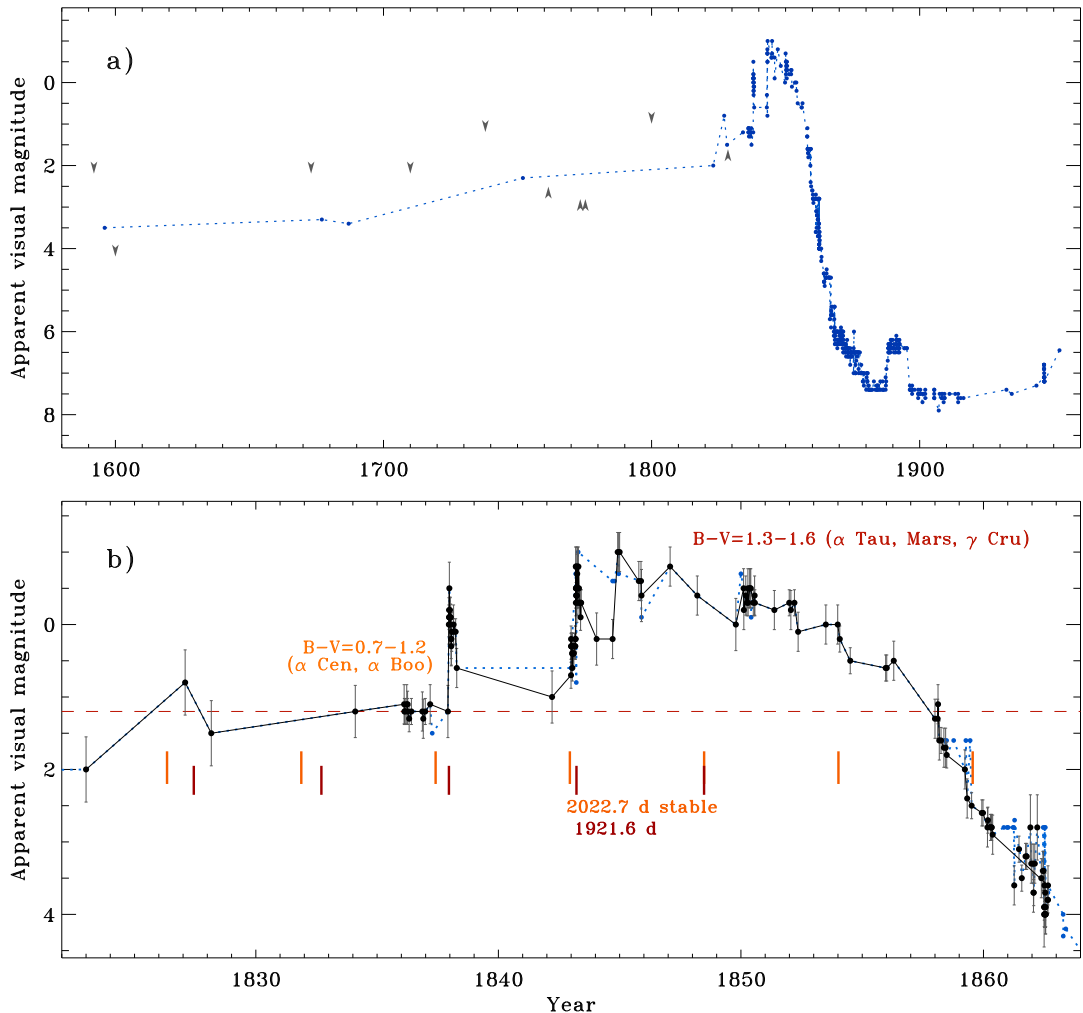}
\end{center}
\caption{The historical light curve for $\eta$ Carinae.  Panel (a)
  shows the full historical light curve from Frew (2004) in blue, with
  limits in gray.  Panel (b) zooms in on the Great Eruption during
  1822--1864.  During this time interval, the previous light curve
  from Frew (2004) is in blue (points and dotted lines), while the
  revised light curve with new archival data that we discuss in this
  paper appears as black dots with error bars.  Notes about the
  apparent color are listed above the light curve.  The orange
  vertical dashes show predicted times of periastron passage if one
  simply extrapolates back from the currently-observed orbital cycle
  with a stable 2022.7 day period (Damineli et al.\ 2008), whereas the
  red hash marks are similar but with a shorter (95\%) period before
  1848.  The dashed red horizontal line shows the quiescent magnitude
  of $\eta$ Car as it would appear with zero bolometric correction.}
  \label{fig:etaLC}
\end{figure*}

\begin{figure*}\begin{center}
\includegraphics[width=5.7in]{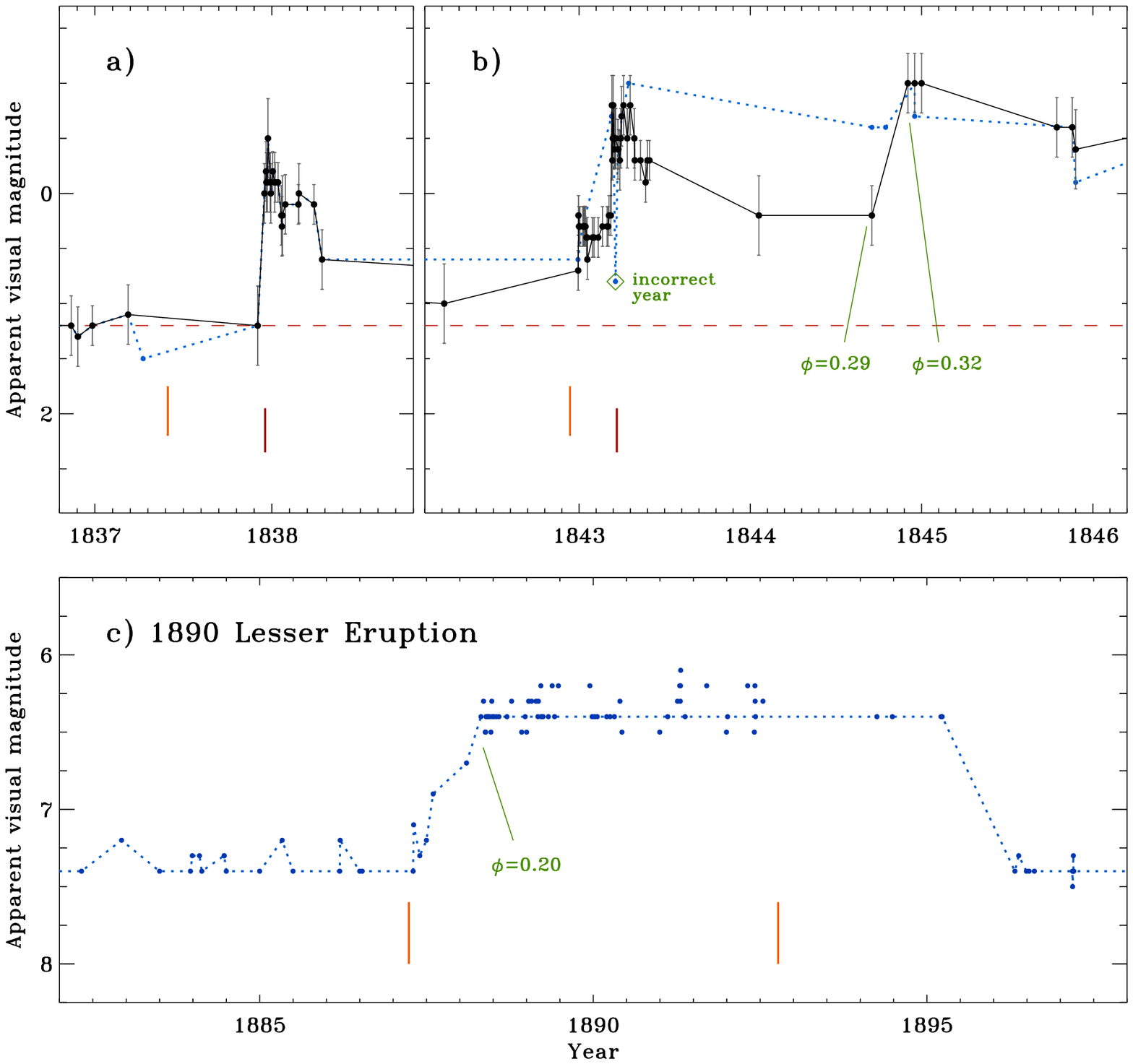}
\end{center}
\caption{Same as Figure~\ref{fig:etaLC}, but zooming in on the events
  in (a) 1837-1838, (b) 1843-1845, and (c) the so-called ``Lesser
  Eruption'' around 1890.}
  \label{fig:etaLCzoom}
\end{figure*}

The qualitative description and large difference in brightness between
Rigel ($\beta$~Ori, $V$ = 0.15) and $\alpha$~Hydrae ($V$ = 1.98)
precludes an accurate estimate for $\eta$~Car in this case.  An
approximate magnitude of 1.0 $\pm$0.5 is inferred, since $\eta$~Car
was somewhat closer in brightness to Rigel. The brightness of
$\alpha$~Crucis sets an upper limit of m$_{V}$ = +0.75.  Since a
brightness `grade' is $\sim$0.2 mag, we again conclude that $\eta$~Car
had m$_{V}$ = 1.0 on this date.  Importantly, this observation
clarifies a discrepancy first noticed by M\"uller \& Hartwig (1918).
Herschel (1847) had mistakenly recorded the wrong date (Mar 19, 1843)
for this observation when compiling his summary of the available data.
The extensive series of observations recorded in the letter by Maclear
(1843) includes no such date.  This error by Herschel has led later
workers to conclude that $\eta$~Car underwent a fast dip and recovery
in early 1843 (e.g., Li et al.\ 2009).  The light curve presented in
Figures~\ref{fig:etaLC} and \ref{fig:etaLCzoom} corrects this error.


\subsection{Results}

Table~\ref{tab:smyth} summarises the new $m_{\rm V}$ magnitudes
derived here.  The columns sequentially list the observer, UT date (as
decimal year), the derived apparent visual magnitude, and the adopted
error on the magnitude.

The first definitive observation of the variability of $\eta$~Car came
from the explorer and naturalist William Burchell in 1827 (see Frew
2004, for a full account of Burchell's observations). Writing from
Brazil on 1827 July 17, Burchell described it as ``now of the first
magnitude, or as large as $\alpha$~Crucis.''  (Herschel 1847).

The star was monitored between 1834 and 1838 by Sir John Herschel at
the Cape of Good Hope (Herschel 1847).  From 1834--37, the star was
essentially constant, with m$_{V}$ = 1.2 $\pm$0.2 (Frew 2004).
Assuming a distance of 2300 pc (Smith 2006), $(m-M)_{0}$ = 11.8 mag,
$A_V$ = 1.4 mag, $M_{\rm bol}$ = $-12.0$ mag, and a bolometric
correction of zero at maximum light (consistent with an F-type
photosphere), we expect an apparent magnitude of m$_{V}$ $\simeq$ 1.2
out of eruption.  The observed brightness before 1838.0 is very
consistent with this estimate (see Figure~\ref{fig:etaLC}).

The Great Eruption is widely considered to have begun at the close of
1838 when Herschel noted a rapid brightnening of $\sim$1 mag over a
period of less than two weeks (Frew 2004).  The star then faded over
the following months but unfortunately we have not recovered any
observations between late 1838 and 1841, so there may have been other
short-duration peaks in brightness that were missed, or the star may
have faded considerably (but see below).  In 1842, the magnitude was
approximately as it was prior to the commencement of the Great
Eruption; our estimate is $m_{\rm V}$ = 1.0 $\pm$0.4 mag.  It was
about 0.5 mag brighter in early 1843 when the brightness suddenly
increased.  The brightness peaked at about $m_{\rm V}$ = $-$0.8
$\pm$0.2 mag in late March 1843.  The star again faded in subsequent
weeks, and for most of 1844 it was constant at $m_{\rm V}$ = 0.2
$\pm$0.2 mag.  At the close of 1844, the star again brightened, and by
January 1845 had reached $m_{\rm V}$ = $-$1.0 $\pm$0.3 mag, which is
brighter than Canopus ($V$ = $-0.74$).

As described by Frew (2004), there is good evidence for marked
fluctuations in brightness (amplitude up to 1 mag on time scales of
days to weeks) during the Great Eruption. The brightening event in
Mar/Apr 1843 was remarkable (see observations of Maclear and
C.P. Smyth described above), as was the brightening at the close of
1844.  After 1846, the observed variations were superposed on a slow
decline (Frew 2004), with fluctuations noted by Jacob (1849) and
Gilliss (1855, 1856).  Between 1846 and 1856, $\eta$~Car faded at an
approximate rate of 0.1~mag\,yr$^{-1}$.  It was still a star of the
first magnitude at the close of 1857, before the rate of fading
suddenly increased by 1859.  This may be due to the onset of dust
condensation from the stellar wind, or the Great Eruption may have
ceased.

Nearly all contemporary reports during the Great Eruption describe
$\eta$~Car as `reddish' or `ruddy' (e.g. Mackay 1843; Smyth 1845;
Jacob 1847; Moesta 1856; Gilliss 1856; Abbott 1861; Tebbutt 1866),
these observers sometimes making direct comparison of its colour with
other stars, or even Mars.  We have estimated an approximate $B-V$
colour index from the these direct comparisons, as summarised in
Table~\ref{tab:eta_color}.  The nominal uncertainties on these
visually estimated colours are approximately $\pm$0.3 mag, following
Schaefer (1996a).  We stress that these values should not be taken as
indictive of the true continuum temperature, as $\eta$~Car probably
had very intense H$\alpha$ emission that would make it appear
considerably redder to the naked eye than its actual (and unknown)
$B-V$ colour index would otherwise indicate.  Nevertheless, the values
in Table~\ref{tab:eta_color} can be used as a \emph{relative
  indicator} to show that Eta tended to redder colours during the
laters stages of the Great Eruption.  This was partly due to a
changing H$\alpha$ equivalent width, but possibly also due to
increasing circumstellar reddening due to dust condensation during the
eruption (note that the grain condensation timescale is roughly 5--10
yr; see Smith 2010).

\begin{table*}\begin{center}\begin{minipage}{3.1in}
      \caption{Colour estimates of $\eta$ Carinae}\scriptsize
\begin{tabular}{@{}lllcl}\hline\hline
Observer        &  Year      & Comparisons & $B-V$  & Reference \\ \hline       
Herschel     &  1834--38     &  = $\alpha$ Cen, $\alpha$ Boo   &  0.9:   &   Herschel (1847)   \\
Herschel     &  1837 Dec 19  &  = $\alpha$ Cen                 &  0.7   &   Evans et al. (1969) \\
Mackay       &  1843 Mar     &  = $\alpha$ Boo                 &  1.2   &   Mackay (1843)    \\
Jacob        &  1845.88      &  redder than $\alpha$ Boo       & $>$1.2 &   Jacob (1847)    \\
Gilliss      &  1850 Feb 9   &  = Mars                         &  1.4:  &   Gilliss (1856)   \\
Gilliss      &  1850 May 28  &  = $\alpha$ Tau                 &  1.6:  &   Gilliss (1856)   \\
Moesta       &  1856 Jan-Aug &  = Mars                         &  1.4:  &   Moesta (1856)     \\
Abbott       &  1858 Mar 6   &  $\gamma$ Cru `somewhat deeper' &  1.3:  &   Abbott (1861)     \\
Abbott       &  1858 Apr 8   &  = $\gamma$ Cru                 &  1.6:  &   Abbott (1861)      \\
\hline
\end{tabular}\label{tab:eta_color}
\end{minipage}\end{center}
\end{table*}

The gap in the light curve between 1838 and 1841 is unfortunate.  Is
it possible that other brief outbursts occurred during this period?
While Maclear and Smyth were at the Cape of Good Hope after Herschel's
departure in 1838, they were occupied by other astronomical pursuits.
However, it is likely they would have noticed if Eta had brightened
beyond zero magnitude, even though they were not to specifically
monitor its brightness until 1842.  Interestingly, the brief outburst
in Mar/Apr 1843 was noticed by three non-professional observers,
specifically Maclean, Leps and Mackay (see Leps 1843; Baily 1843;
Mackay 1843).  Apparently once $\eta$~Car appeared brighter than mag
1, even casual observers noticed it (see also Spreckley 1850).  In
this context, it is germane to mention that indigenous Australians
also appear to have noted $\eta$~Car during its Great Eruption
(Stanbridge 1861; Hamacher \& Frew 2010), incorporating it into their
skylore.  From this we conclude that it seems unlikely that any
significant brightenings between 1838 and 1842 were missed.

Finally, we revisit the observations of Kulczycky (1865), who observed
$\eta$~Car in the 1860s to be brighter than other observers have
recorded (Polcaro \& Viotti 1993).  Feast, Whitelock \& Warner (1994)
and Frew (2004) have cast doubt on the veracity of this report, based
on contemporary data.


\section{Timing of Brightening Episodes}

By modern standards, there is admittedly substantial uncertainty in
the accuracy of reported visual magnitudes of historical accounts.
They are subject not only to atmospheric conditions, transformations
of photometric systems, and variation in the response of the eye from
one observer to the next, but they are subject also to unusually red
colors of $\eta$~Car that may change with time and probably extremely
strong H$\alpha$ line emission.  We have attempted to mitigate these
factors in the historical light curve presented here, and have been
appropriately generous with the uncertainty.

The {\it timing} of relative brightening/fading episodes are quite
reliable, however.  Rare mistakes of transcribed dates in letters
notwithstanding (see above), the timing of reported events are
generally accurate to better than a day.  This provides a powerful
tool to investigate the sequence of events during $\eta$~Car's Great
Eruption, especially as it may pertain to the times of periastron
passage in the putative $\sim$5 year orbit of the binary system.


Damineli (1996) discovered a repeating 5.52 year cycle of
spectroscopic changes in $\eta$~Car that were linked to near-IR
brightening events (Whitelock et al.\ 1994).  Damineli et al.\ (1997)
proposed that these cyclical events were associated with close
periastron passages of a companion star in an eccentric orbit, and the
detailed nature of the orbit and interacting winds has been a topic of
spirited discussion and debate since then.  Damineli (1996) also noted
that three peaks during the Great Eruption seem to coincide roughly
with expected times of periastron, but he only considered the sparsely
sampled data in the light curve of Innes (1903).  The better sampling
in the data presented here and by Frew (2004) allows a closer
investigation of the relative timing of eruptions and periastron
passages.

Figure~\ref{fig:etaLC} shows expected times of periastron passage,
extrapolating back in time from modern events, adopting a period of
2022.7 days and phase 0.0 at year 2003.49 (Damineli et al.\ 2008).
The orange vertical hash marks adopt a stable 2022.7 day period
throughout the Great Eruption.  One can see that expected periastron
passages do not coincide very well with the brief brightening episodes
in 1838 and 1843.  In particular, periastron occurs a few months before
the sharp brightening in 1843 and about 6--7 months before the onset
of the 1838 event; this can be seen more clearly in
Figure~\ref{fig:etaLCzoom}, which conveys the same information but
zooms-in on the time of the individual events.  There may of course be
some slight lag time between the exact time of periastron and the
brightening, depending on exactly how the complicated interaction
occurs physically, but at least {\it we should expect it to be roughly
  the same for both events} if they are related to binary
interactions.  This provides for an unsatisfying link between
periastron passages and brightening events.

A critical point, however, is that the extrapolation above simply
assumed a {\it constant} period throughout the eruption.  This is
certainly invalid.  Observations of the Homunculus indicate that a
very large mass of more than 12.5~$M_{\odot}$ was ejected in the Great
Eruption (Smith et al.\ 2003b).  Smith \& Ferland (2007) note that the
mass could be as high as 20~$M_{\odot}$ but probably not much more, so
$\sim$15 $M_{\odot}$ is a favored value for the mass of the
Homunculus\footnote{Higher estimates of $\sim$40 $M_{\odot}$ based on
  submm emission from cold dust (Gomez et al.\ 2010) include dust
  outside the Homunculus, and possibly free-free emission from ionized
  gas, so 40 $M_{\odot}$ is a generous upper limit to the mass ejected
  in the Great Eruption.}.  As noted in the Introduction, we know that
this mass was ejected during the Great Eruption because of proper
motion measurements of the expanding nebula.  The exact date of origin
for the nebula is still debated; Currie et al.\ (1996) give 1841.2
($\pm$0.8 years), although subsequent authors questioned this date and
the optimistic uncertainty because this study used images taken in
different filters, a short time baseline of only 2 years, and used
abberated pre-COSTAR images with the WFPC camera on {\it HST}.  Smith
\& Gehrz (1998) used a 50-year time baseline and estimated an ejection
date of 1843.8 ($\pm$7 yr), while Morse et al.\ (2001) used corrected
{\it HST}/WFPC2 images with a longer baseline than Currie et al., and
derived dates of 1846--1848 in different imaging filters.
It is not known if the ejection was a sudden singular event (as in a
hydrodynamic explosion) or spread over several years (as in a wind or
multiple ejections).  We consider it likely, however, that the
effective ejection date was around or after the main brightening event
in December 1844, after which $\eta$~Car remained bright for years.
This is only a working hypothesis.  Renewed examination of {\it HST}
images may be worthwhile since the revised light curve we have
presented in this paper raises interesting questions about the exact
time of ejection.

In any case, 15 $M_{\odot}$ is a huge amount of ejected mass.  It is
enough to significantly change the orbit, because mass loss must
reduce the total system mass and gravity, and must therefore make the
period longer after the mass is ejected (that longer post-eruption
orbital period obviously corresponds to the cycle observed in modern
times).  The favored value for the current total stellar mass of the
binary system is $\sim$130 $M_{\odot}$ (assuming a $\sim$100
$M_{\odot}$ primary and a 30 $M_{\odot}$ secondary), which is based on
a number of factors including models of the X-ray light curve and
constraints on the ionizing fluxes and luminosities of the two stars
(see Parkin et al.\ 2009; Okazaki et al.\ 2008; Pittard \& Corcoran
2002; Smith et al.\ 2004; Hillier et al.\ 2001; Corcoran 2005; Mehner
et al.\ 2010).  The ejected nebular mass of $\sim$15 $M_{\odot}$ is
therefore $\sim$11\% of the total remaining stellar mass.  When this
mass was still contained within the star, the gravity was stronger and
the orbital period must have therefore been shorter, at roughly
90--95\% of its present value.

The red hash marks in Figures~\ref{fig:etaLC} and \ref{fig:etaLCzoom}
therefore show times of periastron passage if we reduce the period by
about 5\% before 1844 (to do this we aligned the 1848 periastron
passage with the former value, and used the shorter period before
that).  This shorter period is 1921.6 days (5.26 years). With this
adjusted period, it is quite interesting that the rather sudden
beginnings of the brief 1838 (Figure~\ref{fig:etaLCzoom}a) and 1843
(Figure~\ref{fig:etaLCzoom}b) brightening events both coincide to
within a few weeks with these adjusted times of periastron.  It seems
unlikely that this is a mere coincidence.  There is also a brightening
observed in 1827, which is poorly sampled in time, but is at least
plausibly associated with another periastron passage.

An obvious conjecture, then, is that (somehow) these brief brightening
events are actually {\it triggered} at times of periastron by the
close passage of a companion, as speculated several times before
(e.g., Innes 1914\footnote{Amusingly, the suggestion by Innes that
  ``the outbursts of light which have occurred in the past have been
  caused by periastral grazings'' was based on the first sighting of a
  faint ``companion'' of $\eta$ Car, which is now known to be a very
  distant dust condensation in the equatorial ejecta of the Homunculus
  (see Smith \& Gehrz 1998).  Nevertheless, it illustrates the
  attractiveness of binary systems to explain mysterious
  circumstances.}; Gallagher 1989; Moreno et al.\ 1997 [in the context
of HD~5980]; Iben 1999; Smith et al.\ 2003a; Frew 2004; Kashi \& Soker
2010).  One hypothetical way this might work is if tidal forces from
the close companion push it past a stability threshhold (see Smith et
al.\ 2003a), although the detailed physics of such an encounter have
not been explored.  A more violent encounter may also be possible
(Smith 2011; in prep.).  If true, there should also have been a
similar event in 1831 that was unfortuntely not observed.  A
binary-induced mass ejection could in principle cause the very brief
brightening events if it somehow leads to the ejection of an optically
thick shell.  It seems that theoretical work on the actual effects of
grazing periastron passages that may induce mass ejection in already
unstable stars would be an interesting theoretical pursuit in the
context of eruptive transients.

What happened after 1843? A new result of the present study is that
$\eta$~Carinae faded again after the peak of the 1843 event --- making
this a brief episode akin to the 1838 event --- and it then rose again
to its true peak in December 1844, after which it remained bright.  If
December 1844 really was the beginning of the main phase of the Great
Eruption, then what caused it?  It began suddenly about 1.5 yr {\it
  after} periastron at orbital phase $\phi\approx$0.3, so {\it there
  is no periastron passage that can plausibly be associated with the
  main brightening event}, and subsequent periastron passages do not
appear to induce comparable disturbances during the the rest of the
Great Eruption.  One must conclude, therefore, that close interactions
with a companion at periastron are not the only mechanism that governs
the physics of the Great Eruption, although these interactions may
have pushed an already unstable star past a critical point.

As first pointed out by Frew (2004), the beginning of the smaller
eruption around 1890 is also close to a periastron passage in 1887
(Figure~\ref{fig:etaLCzoom}c).  (This assumes a stable period.  We do
not expect the period to have been altered by the 1890 eruption, since
the total mass ejected was only of order 0.1--0.2 $M_{\odot}$; Smith
2005.)  Curiously, though, the rise time and duration of the event are
extremely different from the brief 1838 and 1843 episodes, which
lasted only a few months.  The delay between periastron passage and
the time when the star rose to maximum brightness is over 1 year, as
opposed to a few weeks in the previous events.  The maximum of the
1890 eruption is actually achieved when the companion star is near its
apastron distance.  Additionally, the 1890 event lasted for about 9
years at roughly constant brightness (Figure~\ref{fig:etaLCzoom}c),
and there is no indication of a major disturbance during the
periastron event that occurs halfway through this eruption (although
the data are quite sparse at the relevant time).  For some unknown
reason, the primary star did not relax after this event was initiated,
and the 1892 periastron passage apparently had little influence.
Altogether, the stark differences between the 1890 event and the
earlier brief events in the Great Eruption raise doubt that the 1890
event was triggered by the same mechanism at a periastron passage; for
such a model to work, we must understand why the 1890 event exhibited
such different timescales.

Independently, Kashi \& Soker (2010) also investigated the possible
relationship between the timing of periastron passages and brightening
events, although they favored a very different scenario from the one
we have outlined above.  In an extended series of papers, Kashi \&
Soker (2010; and several references therein) have advocated a model
wherein the secondary star accretes material from the primary wind at
periastron, increasing the luminosity through accretion and driving a
pair of jets that shape the Homunculus.  One problem with such a model
in the current context is that accretion of a substantial amount of
mass by the companion will tend to contract the orbit and shorten the
period, such that the orbital period would have been {\it longer}
before the Great Eruption than it is now.  As we have seen above,
however, agreement between times of periastron and brightening events
require the opposite -- that the orbital period was about 5\% {\it
  shorter} prior to the Great Eruption (Figures~\ref{fig:etaLC} and
\ref{fig:etaLCzoom}).  Since mass accretion is expected to occur at
periastron, this would appear to contradict a key prediction of the
accretion model.  To mitigate the shortening of the period due to
accretion, Kashi \& Soker (2010) adjusted the model so that the
primary star ejects enough mass to compensate for the accretion and
thereby makes the period longer instead, as we have suggested above,
but with a much larger amount of mass and gravity involved.  In order
to adjust the period enough to match the timing of brightening events
and periastron passages, the favored model of Kashi \& Soker requires
an ejected mass of 40 $M_{\odot}$, as well as present-day stellar
masses of 200 and 80 $M_{\odot}$ for the primary and secondary,
respectively.  These exceed current observational estimates by factors
of 2--3, and would imply an astonishing initial mass for the primary
star of more than 300 $M_{\odot}$.  It seems more straightforward to
conclude that the evolution of the orbital period is dominated simply
by the mass known to be lost from the system by the primary with
conventional stellar parameters, as we proposed above.

\begin{figure}\begin{center}
\includegraphics[width=3.1in]{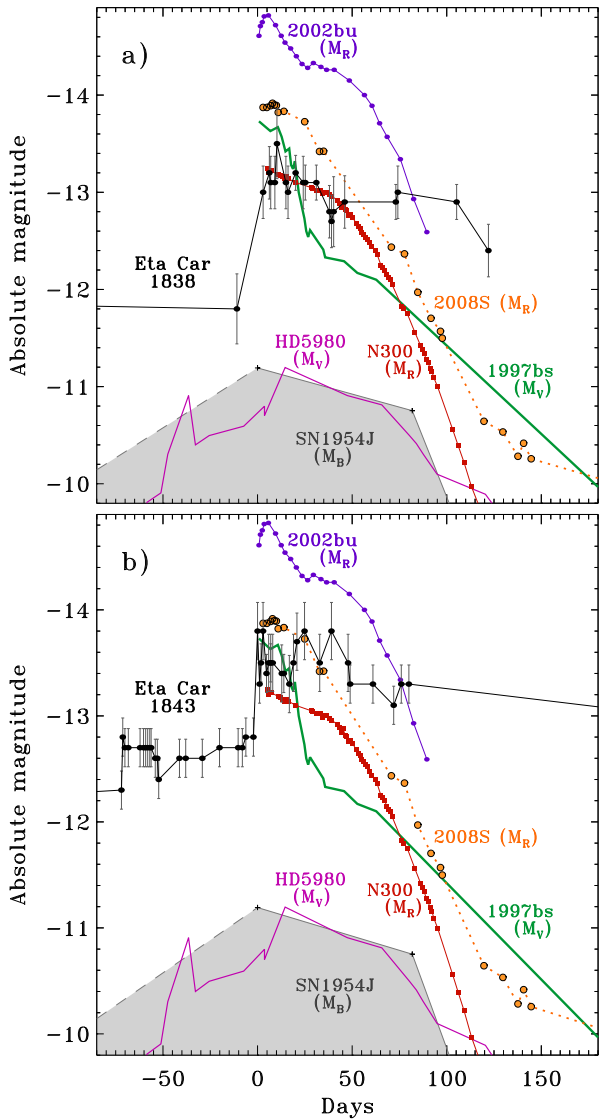}
\end{center}
\caption{A comparison of the revised visual light curves of $\eta$
  Car's brief eruptions in 1838 (a) and 1843 (b) to several SN
  impostors.  This is adapted from Smith et al.\ (2010), where more
  details on each object can be found.  The SN impostors are
  V12/SN~1954J in NGC~2403 (shaded gray; from Tammann \& Sandage
  1968), HD~5980 the SMC (magenta; this is the smoothed version of the
  light curve that appeared in Smith et al. 2010), SN~1997bs (green;
  from Van Dyk et al.\ 2000), SN~2002bu (purple; from Smith et al.\
  2010), SN~2008S (orange; from Smith et al.\ 2009), and the 2008
  transient NGC~300-OT (red; from Bond et al.\ 2009).}
  \label{fig:impostors}
\end{figure}

\section{Qualitative Comparison with Supernova Impostors}

A key result of the new historical magnitude estimates presented here
is that the brightening in 1843 was a brief event, similar to the
precursor brightening in 1838, after which the star faded on a
timescale of a few months before finally surging to its peak at the
end of 1844 when the extended brightening of the Great Eruption began.
Not only is the brief 1843 event similar to the one in 1838, but it
also resembles several examples of so-called ``SN impostors''
discovered in modern times in the course of SN searches.  In the
discussion below, we borrow from a more detailed discussion and
comparison of SN impostors by Smith et al.\ (2010); the reader is
referred to that paper for more details of the general phenomenon.

Figure~\ref{fig:impostors} shows the revised light curves for the
brief eruptions of $\eta$ Car in 1838 (Figure~\ref{fig:impostors}a)
and 1843 (Figure~\ref{fig:impostors}b), shown on an absolute magnitude
scale.  These are compared to the $V$ or $R$ band light curves for
several other SN impostors, taken from Smith et al.\ (2010).

The 1838 eruption has a peak magnitude of $-$13.5, most similar to
N300-OT, and intermediate between SN~2002bu (one of the most luminous)
and SN~1954J or HD~5980.  It appears to fade after 100-120 days.  In
Figure~\ref{fig:impostors}a, we only plot the light curve up to about
day 120, because after that point there are no observations available
until the beginning of the 1843 event several years later, so we do
not know how quickly or how much it faded.  Still, the rate of decline
up to that point appears to be somewhat slower but similar to the
other impostors shown.  The 1843 event had a slightly more luminous
peak magnitude of around $-$13.8, comparable to SN~1997bs or SN~2008S.
It remained luminous for about 80 days, but the behavior after that is
difficult to judge due to a lapse in the observational record.  It
seems likely that a primary difference between these brief precursor
events of $\eta$~Car and the other SN impostors in
Figure~\ref{fig:impostors} is that $\eta$ Car did not fade very much
afterward.  This is probably because it was a more luminous star to
begin with, and also because it was obviously not yet finished
erupting by this point.  The similarity of $\eta$ Car's brief events
to the SN impostors is interesting, and may eventually provide insight
to understand the physical parameters and causes of these
extragalactic events.  So far, two other extragalactic SN impostors
have exhibited repeated eruptions: Pastorello et al.\ (2010) recently
reported that SN~2000ch (LBV1 in NGC~3432; see Wagner et al.\ 2004;
Smith et al.\ 2010) suffered at least three similar brief eruptive
events in 2008 and 2009, and Drake et al.\ (2010) have just recently
reported another outburst of SN~2009ip.  Given that we have noted a
clear connection between the brief eruptions and times of periastron
in the binary system of $\eta$~Car, it is interesting to speculate
that something similar may be occurring in these repeated events in
SN~2000ch/LBV1 and SN~2009ip, and possibly in other SN impostors.
Continued observations may reveal or rule-out true periodicity in the
brightening events.

Following the 1843 event, $\eta$~Car faded to a magnitude that was
somewhat brighter than its expected quiescent magnitude
(Figure~\ref{fig:etaLC}) for about a year.  It then rebrightened
dramatically in December 1844, finally reaching its peak absolute
magnitude at the start of 1845, and remaining luminous for a decade
thereafter.  This behavior is unlike any of the SN impostors shown in
Figure~\ref{fig:impostors}, but there are other SN impostors or LBV
giant eruptions that evolve more slowly and stay bright for years.
Some well-known examples are P Cygni, UGC~2773-OT, and V1 in NGC2366
(see Smith et al.\ 2010 for more details).  We will discuss the
historical light curve of P Cygni in an upcoming paper that is in
preparation. The cause of these longer-duration giant eruptions is
still unknown, and it is not clear if they represent the same
phenomenon as the brief SN impostor events.  Studies of a larger
number of these events over longer time intervals are needed.

\section{Conclusions}

In this paper, we have revisited the historical 19th century light
curve of $\eta$ Carinae, based on 51 newly uncovered historical
estimates of its apparent brightness made at critical times near the
peak of its Great Eruption.  These new estimates correct some previous
mistakes and misconceptions about the light curve, hopefully providing
a definitive historical record, and lead us to several main
conclusions:

1.  The light curve clearly shows two brief ($\sim$100 day) peaks
during the time leading up to the eruption, in 1838 and 1843.  $\eta$
Car then faded by $\sim$1 mag after the 1843 event, before
rebrightening to its true maximum brightness in December 1844.  This
last brightening in late 1844 probably marks the true beginning of the
Great Eruption, which lasted until about 1858 when the star faded
below its quiescent luminosity.

2.  The brief 1838 and 1843 events do not coincide with times of
periastron in the eccentric binary system if we simply extrapolate the
currently observed orbital period back to that time.  However, if the
pre-1844 orbital period is {\it shorter} by $\sim$5\% --- as it should
be due to the considerable mass lost from the system --- then the
peaks of the brief 1838 and 1843 events both occur within weeks of
periastron.  We therefore speculate that these brief brightening
events are somehow triggered at periastron.

3.  A possible brightening may also be associated with an expected
time of periastron in 1827, but the available data are too sparsely
sampled to draw a more firm conclusion.

4.  The final rise to peak in late 1844 occurred at orbital phase
$\phi\approx$0.3, more than 1.5 yr after periastron, so we conclude
that this final event is not triggered by the same mechanism as the
previous brief outbursts.  Similarly, although the {\it beginning} of
the lesser 1890 outburst seemed to occur around periastron, it took
over a year to brighten and reached its maximum brightness when the
system was near apastron, remaining bright for a decade thereafter.
Furthermore, periastron events that should have occurred halfway
through the $\sim$10 yr duration of both the Great Eruption and the
1890 eruption seemed to have little effect.  Thus, periastron
encounters are not likely to be directly responsible for these two
long-duration events.

5.  The light curves of the brief 1838 and 1843 events of $\eta$ Car
are very similar to several other SN impostors.  We speculate that
SN~2000ch, SN~2009ip, and perhaps other brief outbursts may be related
to similar periastron encounters like the 1838 and 1843 eruptions of
$\eta$~Car.  This will be discussed more fully in a separate paper
(Smith 2011, in prep.).

\smallskip\smallskip\smallskip\smallskip
\noindent {\bf ACKNOWLEDGMENTS}
\smallskip
\footnotesize

We thank the staff at Macquarie University Library and the University
of Melbourne Library, Sandra Ricketts (AAO Library).  We are grateful
to B.\ Gaensler for inviting N.S.\ to Sydney and hosting his trip,
providing an opportunity for N.S.\ and D.J.F.\ to meet and discuss
this project.


\end{document}